\documentclass[11pt,twocolumn,twoside]{article}
\usepackage{fully3d}
\usepackage{xcolor}
\usepackage{siunitx}
\usepackage{caption}
\usepackage{subcaption}
\usepackage{booktabs}\usepackage{siunitx}\DeclareSIUnit\kv{kVp}

\begin{document}

\title{ Application of Time Separation Technique to Enhance C-arm CT Dynamic Liver Perfusion Imaging }

\author[1]{Hana Haselji\'{c}}
\author[1]{Vojt\v{e}ch Kulvait}
\author[1]{Robert Frysch}
\author[2]{Bennet Hensen}
\author[2]{Frank Wacker}
\author[1]{Georg Rose}
\author[2]{Thomas Werncke}

\affil[1]{Institute for Medical Engineering and Research Campus STIMULATE, University of Magdeburg, Magdeburg, Germany} 

\affil[2]{Institute of Diagnostic and Interventional Radiology, Hannover Medical School, Hannover, Germany}

\maketitle
\thispagestyle{fancy}

\begin{customabstract}
Perfusion imaging is an interesting new modality for evaluation and assessment of the liver cancer treatment. C-Arm CT provides a possibility to perform perfusion imaging scans intra-operatively for even faster evaluation. The slow speed of the C-Arm CT rotation and the presence of the noise, however, have an impact on the reconstruction and therefore model based approaches have to be applied. In this work we apply the Time separation technique (TST), to denoise data, speed up reconstruction and improve resulting perfusion images. We show on animal experiment data that Dynamic C-Arm CT Liver Perfusion Imaging together with the processing of the data based on the TST provides comparable results to standard CT liver perfusion imaging.
\end{customabstract}

\section{Introduction}

Perfusion CT imaging is an important modality for the treatment of the liver cancer, see \cite{Ogul2014}. It would be the additional benefit to have C-arm CT available as part of the interventional suite, see \cite{Zitzelsberger2018, Hoven2015}. The ability of C-arm systems to measure parenchymal blood volume (PBV) has been the subject of research during the last few years, see \cite{Zitzelsberger2018, Peynircioglu2015, Syha2016, Mueller2016, ODonohoe2019}. The study in \cite{Datta2017} has used animal model to evaluate the dynamic reconstruction algorithm \cite{Manhart2013} and measure Arterial Liver Perfusion using C-arm CT data.

Using C-Arm CT perfusion imaging in the liver cancer management could provide an option to evaluate the performed embolization intraoperatively. The data moreover could be used to plan ablation.  The slow rotation time, limited number of projections and time gap between rotations causes undersampling and artifacts in the contrast agent dynamics reconstruction. Neglecting these limitations by static reconstruction i.e. reconstruction of each rotation individually as if it was a native CT scan, will cause loss of accuracy and could result in incorrect perfusion measurements, see \cite{Fieselmann2013}. To overcome these problems we use the model-based approach to describe time attenuation curves (TAC) as a weighted sum of temporal basis functions, so called Time separation technique (TST), see \cite{Neukirchen, Bannasch2018}.

In this paper we use the data from the C-Arm CT perfusion reconstruction of the swine liver. We are solving the aforementioned problems by using an analytical basis derived from Fourier analysis and applying TST. Then we use the deconvolution based algorithms together with Tikhonov stabilization to compute perfusion parameters, see \cite{Fieselmann2011}. Using this approach we reduce noise in the data and show that produced perfusion maps are comparable to the CT perfusion data.

 \bigskip
\section{Materials and Methods}

\subsection*{Procedure}
We used 2 anaesthetized domestic pigs to perform C-Arm CT perfusion scans of the liver using iodinated contrast agent after the embolization that induced the area of decreased perfusion.  To induce areas of hypoperfusion in the swine liver model we embolized branches of the right hepatic artery with tantalum-based embolization material (Onyx) and coils. 
Two matching C-arm and CT perfusion scans were acquired using Siemens ARTIS pheno C-arm and SOMATOM Force CT. A 15ml of contrast material Imeron 300 was injected with the duration of 5s. The tube voltage was set to \SI{90}{\kv}. Each C-arm scan consisted of five forward-backward sweep pairs. Each sweep covered rotation of \SI{200}{\degree} and with angular step of \SI{0.8}{\degree} acquired 248 projections. 
The scans with the two scanners were performed ten minutes apart to insure the contrast material has washed out.

\subsection *{Time separation technique}
Let the interval $\mathcal{I} = [0,T]$ represent the duration of the scan. The TACs are modeled as a linear combination of a defined set of $N$ orthogonal functions
\begin{equation}
    \mathcal{B} = \{\Psi_1, \ldots, \Psi_N\},
\end{equation}
where for each $i\in\{1, \ldots, N\}$ $\Psi_i = \Psi_i(t), t \in I$ are scalar functions of the time. We call the set $\mathcal{B}$ basis and refer to functions $\Psi_i$ as to the basis functions. In practice these functions might be analytical functions and therefore $I = \mathcal{I}$ or $\Psi_i$ can be represented as a vector of its values in the $M$ time points $I = \{0, T/(M-1), \ldots, T\}$. According the TST, the time attenuation curve in a particular volume point $x_{v}$ is given by the linear combination of the basis functions
\begin{equation}
x_{v}(t) = \sum_{i = 1}^{N} w_{v,i} \psi_i(t). \label{vox}
\end{equation}

Under the assumption of the orthogonality of the basis functions, we can transform the contrast agent dynamic reconstruction problem to the $N$ standard CT reconstruction problems to reconstruct weight coefficients $w_{v,i}$, see \cite{Bannasch2018} for the details. \newpage To do so, we assume that the projection data for any C-Arm spatial configuration, namely angle and pixel position, encoded by index $k$ satisfies
\begin{equation}
p_{k}(t) = \sum_{i = 1}^{N} \omega_{k,i} \psi_i(t). \label{prj}
\end{equation}
Note that by means of equations \eqref{vox} and \eqref{prj} we separate time development, encoded by basis functions $\Psi_i(t)$, from the spatial configuration encoded by weighting coefficients $w$ and $\omega$ respectively, thus the method is called time separation technique.  

In order to reduce noise in the data and extract important information about the contrast agent dynamics, we use trigonometric functions as a basis. 
Based on the numerical experiments we figured out that $N=5$ provides a good tradeof between the number of reconstructions and the image quality. Therefore we have chosen $N=5$ and $\Psi_i$ to be
\begin{equation}
\begin{split}
    \Psi_{0} = 1, \quad \Psi_1&=\sin(\frac{2\pi t}{T}), \quad \Psi_2=\cos(\frac{2 \pi t}{T}),\\ \Psi_3&=\sin(\frac{4 \pi t}{T}), \quad \Psi_4=\cos(\frac{4 \pi t}{T}).
\end{split}
\end{equation}
These functions are orthogonal with respect to the scalar product $<\Psi_i, \Psi_j> = \int_0^T \Psi_i(t) \Psi_j(t)\, \mathrm{d}t$. To find the weighting coefficients of the projection data $w^p$,  we performed a least squares fitting. This was followed by the 40 iterations of the algebraic reconstruction, see \cite{Kulvait2021} \footnote{Source code of the reconstruction technique, namely CGLS, which has been used can be found at 
\url{https://github.com/kulvait/KCT_cbct}.}
and the TAC data was obtained using \eqref{vox}.

In case of CT perfusion data, the acquisition on SIEMENS SOMATOM Force was followed by the analytical reconstruction using Br36 kernel in syngo CT VA50A software. The reconstructed volume data was interpolated by means of cubic splines to obtain TAC data. The estimation of the perfusion maps by our software is identical for both modalities by the method that follows.

\subsection*{Perfusion Parameters Estimation}
The artery input function (AIF) is needed to compute the blood flow through the organ. It describes the contrast agent flow over time. To generate comparable perfusion maps artery was detected as suggested in \cite{Fieselmann2011}. 

We estimate TAC in every voxel as a  convolution of AIF with residual function
\begin{equation}
   \mathrm{tac(t)} = \mathrm{aif(t)} * f_r\mathrm{(t)}.\label{eq} 
\end{equation}
We discretize the function $aif(t)$ by its values in $k=100$ time points. Then we apply the pseudoinverse with Tikhonov regularization to \eqref{eq} in order to recover function $f_r\mathrm{(t)}$.

We compute four perfusion parameters, blood flow (BF), blood volume (BV), mean transit time (MTT) and time to peak (TTP) using the following formulas from \cite{Fieselmann2011}:
\begin{equation}
\begin{split}
        BF = \max{f_r(t)},\quad
        BV =\sum_{i=1}^n f_r\mathrm{(i)},
        \\MTT = \frac{BV}{BF}, \quad
        TTP = \arg\max_{t} f_r(t).
\end{split}
\label{eq:parameters}
\end{equation}

 \label{sec:methods}
\section{Results}

\begin{figure}
     \centering
     \begin{subfigure}{0.41\columnwidth}
         \centering
         \includegraphics[width=\textwidth]{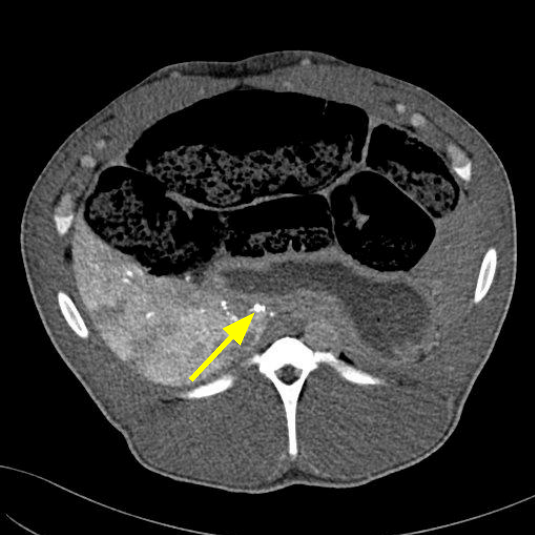}
         \caption{Location of CT AIF}
     \end{subfigure}
\begin{subfigure}{0.55\columnwidth}
         \centering
         \includegraphics[width=\textwidth]{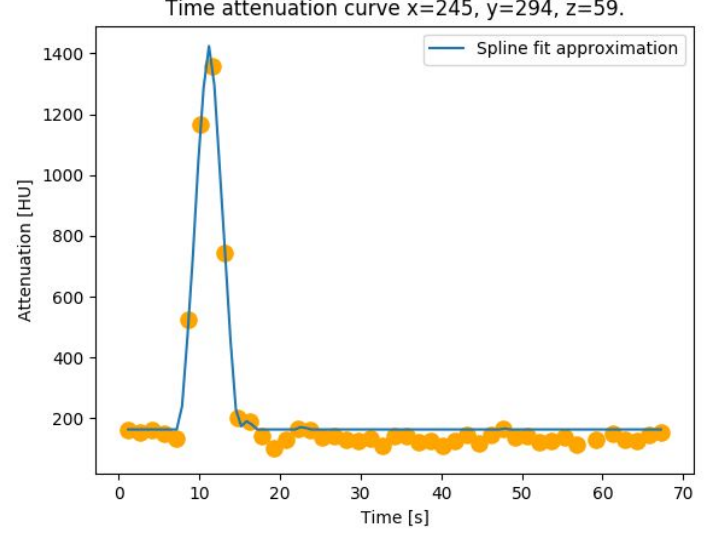}
         \caption{TAC for CT AIF}
     \end{subfigure}
     \vskip\baselineskip
     \begin{subfigure}{0.41\columnwidth}
         \centering
         \includegraphics[width=\textwidth]{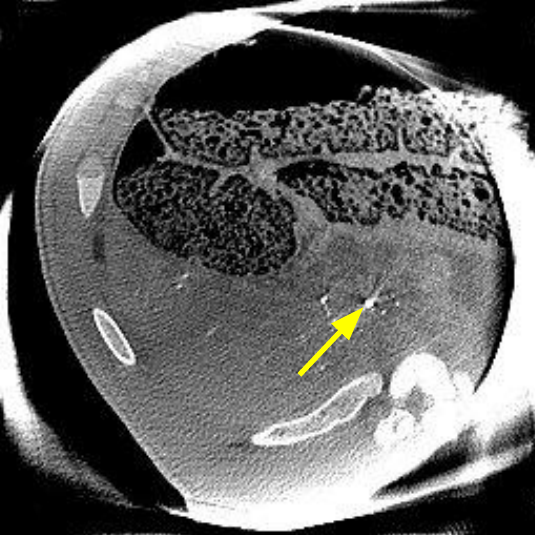}
         \caption{Location of C-arm AIF}
     \end{subfigure}
\begin{subfigure}{0.55\columnwidth}
         \centering
         \includegraphics[width=\textwidth]{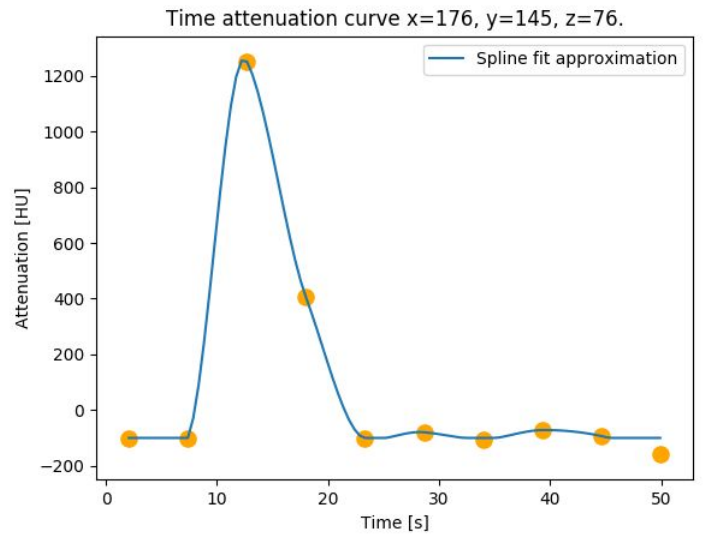}
         \caption{TAC for C-arm AIF}
     \end{subfigure}
        \caption{C-arm and CT artery input function}
        \label{fig:aifs}
\end{figure}

We processed the data from the swine liver perfusion using methods described in M\&M section. To compute perfusion parameters, first we have to locate the arterial inlet in order to derive AIF function. In Figure~\ref{fig:aifs} the detected location of AIF is shown together with the time attenuation profile of this voxel from both modalities. Due to the injection duration, an undersampling of the contrast material peak in C-arm CT compared to CT can occur. 

\begin{figure*}[ht]
  \centering
\includegraphics[width=\textwidth]{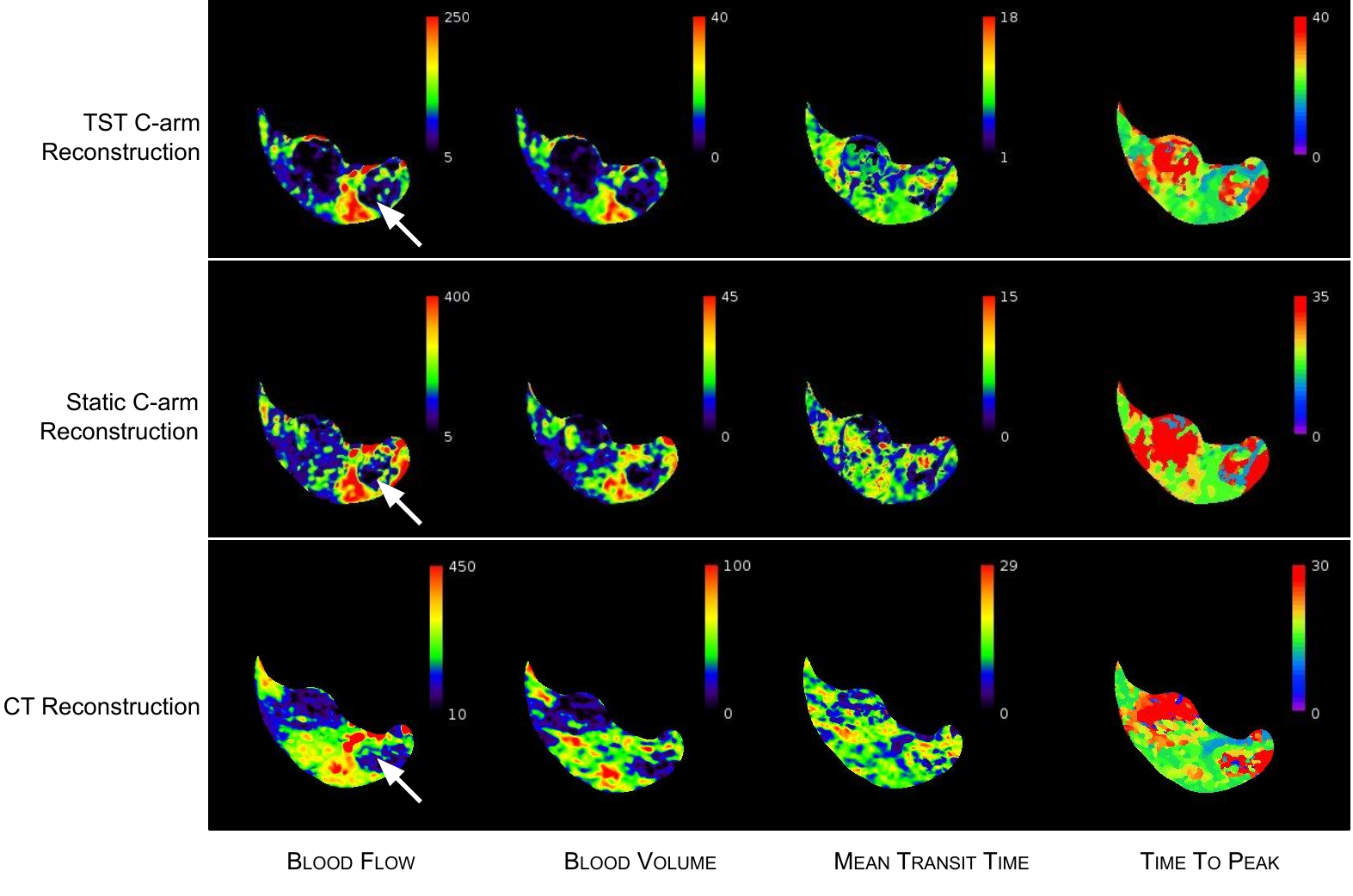}
  \caption{Perfusion maps, BF in \si{\milli\liter\per100\g\per\min}, BV in \si{\milli\liter\per100\g}, MTT in \si{\second} and TTP in \si{\second}, white arrow is pointing to the area of reduced perfusion induced by embolization.}
  \label{fig:perfsmaps}
\end{figure*}

Using the model \eqref{eq} and \eqref{eq:parameters} we have computed perfusion maps. The results for a selected C-arm slice and corresponding CT slice are shown in Figure~\ref{fig:perfsmaps}. The perfusion maps generated by the means of described TST technique are given in the first row, in the second row are the results from the TAC obtained by spline interpolation of the static reconstructions and the third row contains the CT perfusion maps. 

\begin{figure}
  \centering
  \includegraphics[width=\columnwidth]{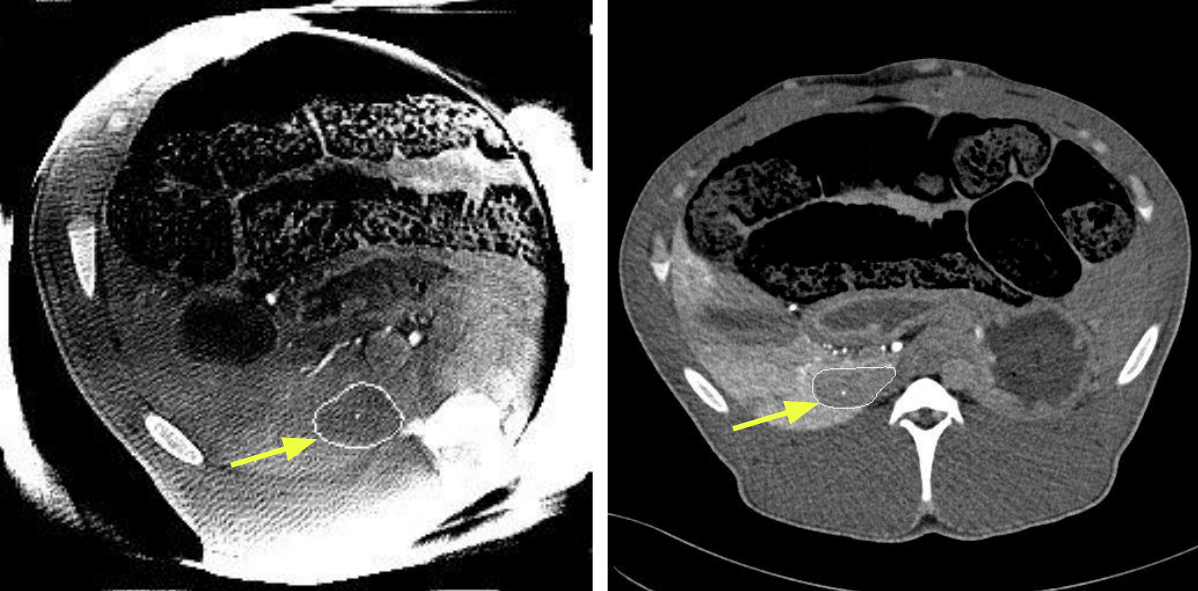}
  \caption{Hypoperfused area in C-arm (left) and CT (right).}
  \label{fig:hypoperfs}
\end{figure}

To compare the results we mainly look for the hypoperfused areas. 
They are easily distinguishable in the reconstruction image, see Figure~\ref{fig:hypoperfs}. 
We can clearly observe these areas in the perfusion maps as they have different color, see Figure~\ref{fig:perfsmaps}. \newpage The unhealthy tissue induced by embolization is very well distinguishable namely in the BF and BV maps of the TST reconstruction, where they have dark blue color code.

We observe that in the TST C-arm reconstruction perfusion maps the hypoperfusion area is much more pronounced than in the static C-arm reconstructions.

\begin{figure}[h]
  \centering
  \includegraphics[width=\columnwidth]{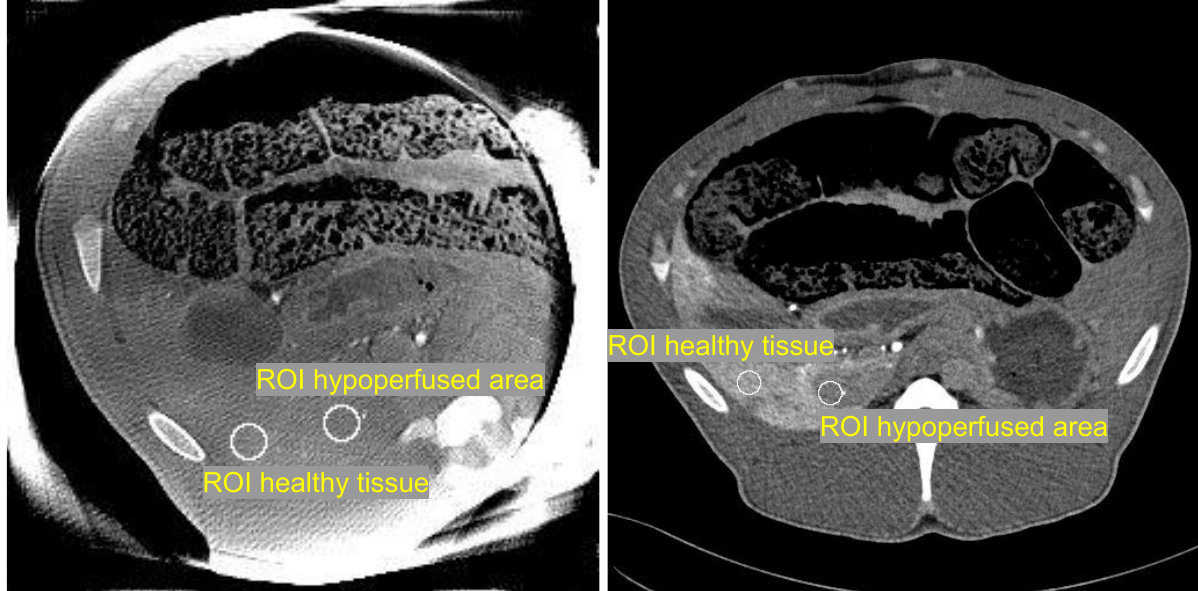}
  \caption{C-arm (left) and CT (right) regions of interest.}
  \label{fig:rois}
\end{figure}

To better evaluate the differences between hypoperfusion area and healthy tissue, we selected regions of interest in both CT and C-Arm CT images, see Figure \ref{fig:rois}. We then computed mean value and standard deviation of the respective perfusion coefficient.

\newpage We have found that not only the means are different, but when performed T-test to find out the significance of that difference, we found a p-value of < $10^{-10}$. The mean value and standard deviation of BF and BV for selected regions are given in Tables \ref{tab:bfmeasure} and \ref{tab:bvmeasure}. 

\sisetup{separate-uncertainty}
\begin{table}[h]
 \centering
 \begin{tabular}{cS[table-format=-2.2(3)]r@{\,\( \pm \)\,}l}
 \toprule
 Reconstruction   & \textrm{Hypoperfused area} & \multicolumn{2}{c}{Healthy tissue} \\
 \midrule
 TST        & \num{20.7(204)}    &\num{95.6} & \num{24.3}\\
Static      & \num{38.8(396)}     &  \num{133.4} & \num{41.4} \\
 CT         & \num{91.0(329)}    & \num{1162}& \num{122} \\
 \bottomrule
 \end{tabular}
 \caption{BF measurements in units of \si{\milli\liter\per100\g\per\min}. Mean and the standard deviation taken over the selected regions of interest.}
 \label{tab:bfmeasure}
\end{table}

\sisetup{separate-uncertainty}
\begin{table}[h]
 \centering
 \begin{tabular}{cr@{\,\( \pm \)\,}lr@{\,\( \pm \)\,}l}
 \toprule
 Reconstruction & \multicolumn{2}{c}{Hypoperfused area} & \multicolumn{2}{c}{Healthy tissue} \\
 \midrule
 TST   & \num{2.3} & \num{3.0} & \num{13.4} &  \num{3.1} \\
Static & \num{1.9} & \num{2.4} & \num{19.1} &  \num{5.0} \\
 CT    & \num{18.8} & \num{8.6}  & \num{58.9} & \num{14.1} \\
 \bottomrule
 \end{tabular}
 \caption{BV measurements in units of \si{\milli\liter\per100\g}. Mean and the standard deviation taken over the selected regions of interest.}
 \label{tab:bvmeasure}
\end{table}

 \section{Discussion}

In this paper we have shown that C-Arm CT liver perfusion imaging can provide similar results as CT perfusion imaging when we use an adequate model based approach. \newpage It can be seen that proposed basis set of the TST provides a perfusion maps that reduce the noise and clearly separates the hypoperfusion areas from healthy tissues, see Figure~\ref{fig:perfsmaps} and Tables \ref{tab:bfmeasure} and \ref{tab:bvmeasure}. Thus, this method enables detection of hypoperfusion regions and has the potential to be introduced into clinical practice.

C-arm perfusion maps show also some differences in well perfused areas when compared to CT perfusion maps. The difference in values are mostly noticeable on BF and BV. This can be mainly attributed to the fact that the positioning of the animal in two different modalities is different but the different setup of the reconstruction for the two modalities can also play its role, see \cite{Mueller2016}. C-Arm and CT devices also were not calibrated to provide equal attenuation values.

 \section{Conclusion}

From the results it can be seen that model based reconstruction of the C-Arm CT perfusion scans outperforms methods based on individual static reconstructions. The data are less noisy and the area with reduced perfusion is more visible on perfusion maps. Therefore it provides the results comparable to the CT perfusion imaging. We plan to perform clinical evaluation of these data to assess whether this approach should be part of the clinical setup.

Additional improvement of the TST results is expected by including dedicated perfusion basis functions based on CT data as prior knowledge, as studied for brain data in \cite{Eckel2018, Bannasch2018}.

\section*{Acknowledgments}
\textit{This work was partly funded by the European Structural and Investment Funds (International Graduate School MEMoRIAL, project no. ZS/2016/08/80646) and the German Ministry of Education and Research (Research Campus STIMULATE, grant no. 13GW0473A and 13GW0473B).} 

\printbibliography

\end{document}